
\magnification=\magstep1
\parindent=0pt
\parskip=6pt
\overfullrule=0pt
\input amssym.def
\input amssym.tex

\def\today{\ifcase\month\or
  January\or February\or March\or April\or May\or June\or
  July\or August\or September\or October\or november\or December\fi

  \space\number\day, \number\year}
\font\largebf = cmbx12 scaled \magstep1
\hsize 6.25 true in
\vsize 9.25 true in
\openup 1 \jot
\rightline {DAMTP 95-29}
\vskip 2.5cm
\centerline {\largebf The Moduli Space Metric for Well-separated}
\centerline {\largebf BPS Monopoles}
\vskip 1.2cm
\centerline {\bf G.W. GIBBONS \& N.S. MANTON}
\vskip 1.2cm
\centerline {Department of Applied Mathematics and Theoretical Physics}
\centerline {University of Cambridge}
\centerline {Silver Street, Cambridge CB3 9EW, U.K.}
\vskip .25cm
\centerline {\rm \today}
\vskip 3.5cm
\leftline  {\bf {Abstract}}
The Lagrangian for the motion of $n$ well-separated BPS monopoles is
calculated, by treating the monopoles as point particles with
magnetic, electric and scalar charges. It can be
reinterpreted as the Lagrangian for geodesic motion on the asymptotic region of
the $n$-monopole moduli space, thereby determining the asymptotic
metric on the moduli space. The metric is hyperk\"ahler, and is an
explicit example of a type of metric considered previously.
\vskip 5cm
The moduli space for $n$ Bogomolny-Prasad-Sommerfield (BPS) monopoles,
$M_n$, is $4n$-dimensional, and has a natural hyperk\"ahler metric [1].
It has the form $M_n$ =
$\Bbb R^3 \times ( S^1 \times {\tilde M_n^0}) / \Bbb Z_n$,
where ${\tilde M_n^0}$ is simply
connected and is the moduli space of ``strongly centred'' monopoles [2].
The $n$-fold cover of $M_n$, $\ \Bbb R^3 \times  S^1 \times {\tilde
M_n^0}$, is
metrically a direct product, with
$\Bbb R^3 \times S^1$ being flat. The geodesics on $M_n$ accurately
model low energy monopole dynamics [3,4].

It is generally understood that when $n$ monopoles are well-separated in
space $(\Bbb R^3)$, the $4n$ coordinates on $M_n$ can be thought of
as positions and phases of $n$ indistinguishable particles.
Bielawski has shown
that to a first approximation, and working locally, the metric is flat
in these coordinates, and he has identified the coordinates with the
parameters occuring in the rational maps which are associated with
monopoles [5].  However, the detailed topology of this outer region of
moduli space has not been clearly elucidated, neither is the metric
known accurately.  It is known that if $n > 1$, it is not just
the product of a configuration space of $n$ point particles by an
$n$-torus. In this paper we shall calculate the metric on this outer
region of moduli space, and make some remarks on the topology.

In the simplest case, $n=1$, the moduli space $M_1 = {\Bbb R}^3 \times
S^1$ and the metric is flat.  The geodesic motion is at
constant velocity in ${\Bbb R}^3$, and constant angular velocity
$\dot{\theta}$ on the $S^1$-factor.  Physically this describes a
monopole with an electric charge (a dyon) in uniform motion; its
magnetic charge $g$ is fixed and its electric charge is proportional
to $\dot{\theta}$. The geodesic motion on $M_n$, according to
Bielawski's metric, describes the independent uniform motion of $n$
dyons in ${\Bbb R}^3$, each with magnetic charge $g$ and a constant
electric charge.

Now, the true motion of monopoles, even if they are well-separated, is
more interesting.
The $2$-monopole moduli space $M_2$ is known in detail, and its metric has
been determined by Atiyah and Hitchin [6].  The asymptotic metric on
$\tilde M_2^0$, when the two monopoles are well-separated,
is obtained by neglecting all terms in the Atiyah-Hitchin metric which
are exponentially small, and it has a simple algebraic form. It is the
self-dual euclidean Taub-NUT metric [7] with a negative mass parameter.
Geodesics
on Taub-NUT space describe two monopoles or two dyons interacting via
Coulomb-like forces.

The Taub-NUT metric has a $U(1)$ symmetry not possessed by the
Atiyah-Hitchin metric.  This implies that well-separated dyons
have individually conserved electric charges.  In contrast, the
general geodesic motion on $M_2$ allows for electric charge exchange
(as well as momentum exchange) in a close encounter of two monopoles
or dyons.

Shortly after Atiyah and Hitchin obtained their metric, one of the
present authors showed that its asymptotic form can be obtained
from a physical calculation of the dynamics of two well-separated
dyons [8].  It suffices to consider the dyons as point particles,
each with a magnetic charge, electric charge and scalar charge.  The
equations of motion for the two dyons, assuming that
their speeds are modest and that their electric charges are
much less than their magnetic charges,
are found to be equivalent to the equations for geodesic motion
on Taub-NUT space, together with a centre of mass motion. The
scalar charge can be understood from the exact solution of the field
equations for a single dyon.  For a magnetic charge $g$ and electric charge
$q$, the
Higgs field far from the dyon centre has magnitude
$$
v - {(g^2 + q^2)^{1\over 2} \over {4\pi r}} + O (e^{-8\pi vr/g}), \eqno(1)
$$
where $v$ is the vacuum expectation value of the Higgs field.
This may be interpreted as saying that the dyon has scalar charge
$(g^2 + q^2)^{1\over 2}$.

In this paper we shall repeat this calculation, but for $n$ dyons.
Our strategy is to find the Lagrangian describing the
dyons' motion in ${\Bbb R}^3$, with the (constant) electric
charges as parameters, and then to reinterpret the $n$
electric charges as arising from motion on $n$ circles associated with
the $n$ monopoles. In this way, the Lagrangian is shown to be
equivalent to the Lagrangian for geodesic motion on a particular
$n$-torus bundle over the configuration space of $n$ monopole
positions, with respect to a metric whose explicit form we give.

The metric we obtain has some beautiful features, and its explicit
form is quite simple.  It admits an isometric $n$-torus action, as
well as the action of the Euclidean group on ${\Bbb R}^3$.  The
equations of motion for the dyons, to which it gives rise, are
Galilean invariant, which is not necessarily the case for a general moduli
space Lagrangian.  Finally it is hyperk\"ahler.
Hyperk\"ahler metrics on $4n$-dimensional spaces have been studied
previously.  A number of authors have presented formulae for such
metrics involving matrices subject to some differential constraints.
Our metric is in the class previously considered, and one may easily
show that it satisfies the required constraints.

Consider $n$ dyons,
all with the same magnetic charge $g$.  Let the $i$th dyon have
electric charge $q_i$ and scalar charge $(g^2 + q^2_i)^{1\over 2}$,
which we shall later approximate by $g + {q^2_i \over 2g}$;
its rest mass is $m_i = v(g^2 + q^2_i)^{1\over 2}$. Suppose the dyons
have positions ${\bf x}_i$  and velocities ${\bf v}_i$. Denote their
separations ${\bf x}_j - {\bf x}_i $ by
${\bf r}_{ji}$, and set $r_{ji} = |{\bf r}_{ji}|$.
The Lagrangian for the motion of the $n$th dyon in the background
of dyons $ 1, \dots, n-1$ is
$$
\eqalign{
L_n = \Bigl( -m_n &+ (g^2 + q_n^2)^{1\over 2} \phi \Bigr )
(1 - {\bf v}^2_n)^{1\over2}\cr
&+ q_n {\bf v}_n \cdot {\bf A} - q_{n} A_0 \cr
&+ g {\bf v}_n \cdot \tilde{\bf A} - g {\tilde A}_0. \cr} \eqno(2)
$$
Here $\phi$ is the scalar field due to dyons $1, \dots, n-1$; its
effect is to modify the rest mass of dyon $n$, with the
coefficient $(g^2 + q_n^2)^{1\over 2}$ being the scalar charge of dyon
$n$.  ${\bf A}, A_0$ are the vector and scalar Maxwell potentials
due to all but the $n$th dyon, and they couple to the electric charge of
the $n$th dyon, $q_n$.  The magnetic charge of the $n$th dyon, $g$,
couples to the dual Maxwell vector and scalar potentials ${\tilde{\bf
A}, \tilde{A}_0}$ produced by dyons $1, \dots, n-1$.  These potentials
are defined so that
$$
\eqalign {
\nabla \times {\bf A} &= {\bf B} \cr
- \nabla A_0 - \dot{\bf A} &= {\bf E} \cr
\nabla \times { \tilde {\bf A}} &= - {\bf E} \cr
- \nabla {\tilde {A}_0} - \dot {\tilde {\bf A}}  &= {\bf B},\cr
} \eqno (3)
$$
where ${\bf E}$ and ${\bf B}$ are the electric and magnetic fields at
the $n$th dyon, due to the other dyons.

The fields $\phi, {\bf A}, A_0, \tilde{\bf A}$ and $\tilde{A_0}$ are
linear combinations of the fields produced by the $n-1$ contributing
dyons.  The contribution from dyon 1 is given in  terms of the
Li\'enard-Wiechert potentials at ${\bf x}_n$ due to the moving particle
at ${\bf x}_1$.  The scalar field
due to dyon $1$ at ${\bf x}_{n}$ is
$$
\phi = { (g^2 + q^2_1)^{1\over 2} \over 4\pi s_{n1} } (1 - {\bf
v}_1^2)^{1\over 2}.
\eqno(4)
$$
Eqn. (4) is the Lorentz scalar version of a Li\'enard-Wiechert
potential, where
$s_{n1} = \left( r_{n1}^2 - ({\bf r}_{n1} \times {\bf v}_1) ^2
+ o ({\bf v}_1^2) \right )^{1\over 2}$.  Now
the leading terms in the Lagrangian will turn out to be of order
${\bf v} ^2/r$, and this is the order in ${\bf v}$ to which we wish to
work, so it is sufficient to approximate $s_{n1}$ by $r_{n1}$.
Expanding out the square roots, and keeping terms quadratic in
velocities and electric charges, we find
$$
\phi = {g \over 4\pi r_{n1} } \left ( 1 + {q_1^2 \over 2g} - {{\bf v}^2_1 \over
2} \right ). \eqno (5)
$$

To write down the Maxwell and dual Maxwell potentials, we introduce a
Dirac vector potential ${\bf w}({\bf y})$.  This is defined to satisfy
$$
{\bf \nabla} \times {\bf w} = - {{\bf y} \over y^3}
\ \ , \eqno(6)
$$
and ${\bf w}({\bf y}) = {\bf w}(-{\bf y})$, but
we do not need to specify a particular gauge. We denote by
${\bf w}_{ji}$ the potential ${\bf w}( {\bf x}_j - {\bf x}_i) =
{\bf w}({\bf r}_{ji})$; it has the symmetry property
${\bf w}_{ji} = {\bf w}_{ij}$ and satisfies
${\bf \nabla}_j \times {\bf w}_{ji} = {\bf \nabla}_j (1/r_{ji})$.
The potentials ${\bf A}, A_0, \tilde{\bf A}$ and $\tilde{A}_0$
produced by dyon 1, and evaluated at ${\bf x}_n$ are
$$
\eqalignno{
{\bf A} &= {q_1 \over 4\pi r _{n1}} {\bf v}_1 - {g \over 4\pi} {\bf
w }_{n1}&(7)\cr
A_0 &= {q_1 \over 4\pi r_{n1}} - {g \over 4\pi} {\bf w }_{n1} \cdot {\bf v}_1
&(8)\cr}
$$
$$
\eqalignno{
\tilde{\bf A} &= {g \over 4\pi r_{n1}} {\bf v}_1 + {q_1 \over 4\pi} {\bf
w}_{n1}
&(9) \cr
\tilde{A}_{0} &= {g \over 4\pi r_{n1}} + {q_1 \over 4\pi} {\bf w}_{n1}\cdot
{\bf v}_1 . &(10)\cr }
$$
These are boosted versions of the familiar Coulomb and Dirac
potentials of a dyon at rest, and we have again approximated $s_{n1}$
by $r_{n1}$.
Substituting (5) and (7) -- (10) in (2), and keeping terms of order
${\bf v}^2$,
$q{\bf v}$ and $q^2$, we obtain
$$
\eqalign{
L_n &= -m_n + {1\over 2} m_n{\bf v}_n ^2 - {g^2 \over 8\pi r_{n1}}
({\bf v}_n - {\bf v}_1) ^2 \cr
&\qquad - {g\over 4\pi} (q_n - q_1) ({\bf v}_n - {\bf v}_1 ) \cdot
{\bf w}_{n1} + {1\over 8\pi r _{n1}} (q_n - q_1)^2. \cr}\eqno (11)
$$

Eqn. (11) includes the contribution from dyon 1 to the Lagrangian for dyon
$n$.  Adding the contributions of all dyons $1, \dots n-1$ gives
$$
\eqalign{
L_n =  {1\over 2} m{\bf v}_n^2 - {g^2 \over 8\pi}
\sum^{n-1}_{i=1} {({\bf v}_n - {\bf v}_i) ^2 \over r_{ni}}
&- {g\over 4\pi} \sum^{n-1}_{i=1} (q_n - q_i) ({\bf v}_n - {\bf v}_i )
\cdot {\bf w}_{ni} \cr
&+ {1\over 8\pi} \sum^{n-1}_{i=1} {(q_n - q_i)^2 \over r_{ni}}.
\cr} \eqno (12)
$$
(We have dropped the constant $-m_n$, which has no effect on the
equation of motion, and in the kinetic term replaced $m_n$ by $m =
vg$, which is accurate enough to the order we are working.)
It can be seen that the interaction terms are quite symmetric
between dyons $i$ and $n$.
It follows, that if we symmetrize (12) between all the
dyons, we will obtain a Lagrangian whose Euler-Lagrange equations are
the equations of motion for all the dyons.  The symmetrized
Lagrangian is
$$
\eqalign {L = \sum ^{n}_{i=1} {1 \over 2} m {\bf v}_i ^2 -
{g^2 \over 8\pi}
\sum_{1\le i <j  \leq n} { ({\bf v}_j-{\bf v}_i ) ^2 \over r_{ji} }
&- { g \over 4 \pi} \sum _{1\le i<j\le n} (q_j-q_i)({\bf v}_j - {\bf v}_i )
\cdot {\bf w }_{ji} \cr
&+ { 1 \over 8 \pi} \sum _{1\le i <j \le n} { (q_j-q_i)^2 \over r_{ji} }.
\cr} \eqno (13)
$$
(Here we have used the symmetry property ${\bf w}_{ji} = {\bf w}_{ij}$.)
The kinetic term may be rewritten as
$$
\sum^{n}_{i=1} {1\over 2} m {\bf v}_i ^2 = {1\over 2n} m ({\bf
v}_1 + \dots + {\bf v}_n ) ^2 +  \sum_{1\leq i< j \leq
n} { 1\over 2n} m ({\bf v}_{j} - {\bf v}_{i})^2,
\eqno(14)
$$
which shows that the sum of the dyon velocities completely decouples
from the relative velocities.  It follows that ${1\over n} ({\bf v}_1 +
\dots + {\bf v}_n)$ is a conserved quantity; it can be identified as
the centre of mass velocity.  Furthermore, the relative motion is
unaffected by a Galilean transformation ${\bf v}_i \rightarrow {\bf
v}_i + {\bf V}$, for any fixed ${\bf V}$.
There is also invariance under the Euclidean group of translations and
rotations.

The equation of motion for dyon $j$ is
$$
\eqalign {
m \dot{{\bf v}} _j - {g^2\over 4\pi} \sum_{i\neq j}
{\dot{\bf v}_j -\dot{\bf v}_i \over r_{ji} }
+ {g^2 \over 8\pi} \sum_{i\neq j}
{ 2 ({\bf v}_j - {\bf v}_i) {\bf r}_{ji} \cdot ({\bf v}_j - {\bf v}_i)
-  ({\bf v}_j - {\bf v}_i)^2 {\bf r} _{ji}
\over r^3_{ji} } \cr
- {g \over 4\pi} \sum_{i\neq j}  (q_j - q_i) ({\bf v}_j - {\bf v}_i)
\times {{ \bf r} _{ji} \over r^3_{ji}}
+ {1\over 8 \pi } \sum_{i\neq j} {(q_j - q_i)^2 \over r^3_{ji}} {\bf r}_{ji} =
0. \cr }\eqno (15)
$$
The last term in (15), an electric Coulomb force, depends on the
square of the difference of electric charges rather than on their
products.  This is one  of the consequences of the scalar
interaction.  Note that there is a static solution of
the equations of
motion $({\bf v}_i = 0 : 1 \leq i \leq n)$ with all electric charges
equal.  This is to be expected, as there are solutions of the
Bogomolny equations representing $n$ monopoles at rest, and these can be
transformed in a simple way (discovered by Julia and Zee [9]) into a
stationary solution representing $n$ dyons each with the same electric charge.

The Lagrangian (13) is defined on the $3n$-dimensional configuration
space $\tilde{C}_n ({\Bbb R}^3)
= ({\Bbb R}^3)^n - \triangle $ with coordinates $\{ {\bf x}_i : \ 1 \leq
i \leq n \}$, where $\triangle$ represents the subspace of $({\Bbb
R}^3)^n$ where two or more monopole positions coincide. The electric
charges $q_i$ are constant parameters.  The Lagrangian is not purely kinetic,
because of the terms linear in velocity (the electric-magnetic
couplings), and because of the Coulomb terms.  But if we could interpret
the electric charges as velocities, as in Kaluza-Klein theory, then
the whole Lagrangian would be quadratic in velocities.  We shall now
show that this is indeed possible.

Consider a $4n$-dimensional manifold $E_n$, which is a $T^n$ ($n$-torus)
bundle over $\tilde{C}_n ({\Bbb R}^3)$ with local coordinates $\{ {\bf x}_i,
\theta _i \}$. $\theta_i$ is a phase angle associated with the $i$th
monopole.  Endow this
bundle $E_n$ with a $T^{n}$-invariant metric, so that the purely kinetic
Lagrangian for motion on $E_n$ is
$$
{\cal L}  = {1\over 2} g_{ij} {\bf v}_i \cdot {\bf v}_j + {1\over 2} h_{ij}
(\dot{\theta }_i + {\bf
W} _{ik} \cdot {\bf v}_k)
(\dot {\theta} _{j} + {\bf W} _{jl} \cdot {\bf v}_l ), \eqno (16)
$$
where $g_{ij}$ and $h_{ij}$ are symmetric matrices. (From now on we
shall be using the summation convention unless we say to the contrary.)
Invariance under the torus action requires that the matrices $g_{ij},
h_{ij}$ and ${\bf W}_{ij}$ depend only on the $3n$ coordinates $\{
{\bf x}_i \}$.  The Euler-Lagrange equations for $\cal L$ admit the
$n$ constants of motion
$$
q_i = \kappa h_{ij} ({\dot \theta} _j + {\bf W}_{jl} \cdot {\bf v}_l). \eqno
(17)
$$
For a suitable value of the constant $\kappa$, $q_i$ may be identified
with the electric charge of monopole $i$.

It would not be correct to substitute for these constants of motion in
the second term of $\cal L$, and then calculate the remaining equations
of motion.  However, it is correct to reduce the Lagrangian $\cal L$ to
$$
{\cal L} _{\rm eff} = {1\over 2} g_{ij} {\bf v}_i \cdot {\bf v}_j + {1\over
\kappa} q_i {\bf W}_{ij} \cdot{\bf v}_j - {1\over 2\kappa^2}
h^{ij} q_i q_j \eqno (18)
$$
where $h^{ij}$ is the inverse of $h_{ij}$.  The equations of motion
given by $\cal L$ and ${\cal L} _{\rm eff}$ are the
same.  It follows that we can determine $g_{ij},
h_{ij}$ and ${\bf W}_{ij}$ by requiring that
${\cal L}_{\rm eff}$ and the Lagrangian $L$, eqn. (13), give the same
equations of motion. This almost means the same thing as identifying
${\cal L}_{\rm eff}$ and $L$. The matrix $g_{ij}$ must  be
chosen to have components
$$
\eqalignno{
g_{jj} &= m - {g^2 \over 4\pi} \sum_{i \ne j} { 1 \over {r_{ij}} }
\quad {\rm (no~ sum~ over~} j) &(19) \cr
g_{ij} &= {g^2 \over 4\pi} {1\over r_{ij}} \quad (i\ne j), &(20) \cr}
$$
and ${\bf W} _{ij}$ must have components
$$
\eqalignno{
{\bf W} _{jj} &= -{g\kappa \over 4\pi}\sum_{i\neq j} {\bf w}_{ij} \quad ({\rm
no ~ sum ~ over ~}j) &(21) \cr
{\bf W }_{ij} &=  {g\kappa \over 4\pi}{\bf w }_{ij} \quad (i\ne j). &(22)
\cr}
$$
The symmetry properties of the Dirac potentials imply that ${\bf W}_{ij}$
is a symmetric matrix. Simply identifying ${\cal L}_{\rm eff}$ and $L$
would give a matrix $h^{ij}$ with no inverse. But we may add any
constant matrix to $h^{ij}$ without changing the equations of motion
coming from ${\cal L}_{\rm eff}$. Taking advantage of this we see that
a satisfactory choice is
$h^{ij} = {\kappa ^2 \over g^2} g_{ij}$.

The value of $\kappa$ can now be fixed. The expressions (21) and (22)
for ${\bf W}_{ij}$ could lead to singularities in the Lagrangian (16),
because the Dirac potentials have the usual Dirac string
singularities. There are two Dirac strings in the potential
${\bf w}({\bf y})$, because of the symmetry of the potential, and each
carries ``flux'' $2\pi$. These strings can be gauged away if $\kappa =
4\pi /g$, and if the angles $\theta _i$ have the usual range
$[0,2\pi]$.

To obtain the Lagrangian $\cal L$ in its simplest mathematical form, it
is convenient to choose units so that $g=4\pi, m=4\pi$ (these are in
any case true for BPS monopoles in the natural units [1])
and to remove an overall factor of $4 \pi$. $\cal L$ is
the purely kinetic Lagrangian for motion on $E_n$ with metric
$$
ds^2 = g_{ij} d{\bf x}_i \cdot d{\bf x}_{j} + g^{-1}_{ij} (d\theta_i + {\bf
W}_{ik} \cdot d{\bf x}_k)( d\theta_{j} + {\bf W}_{jl} \cdot
d{\bf x}_l) \eqno (23)
$$
where, now,
$$
\eqalignno{
g_{jj} &= 1 - \sum_ {i \ne j}{1\over r_{ij}} \quad  ({\rm no~ sum ~over ~}j)
&(24)\cr
g_{ij} &= {1\over r_{ij}} \quad (i \neq j)&(25)\cr
{\bf W}_{jj} &=  - \sum_{i \ne j}  {\bf w}_{ij}  \quad  ({\rm no~ sum ~over
{}~}j) &(26)\cr
{\bf W}_{ij}  &= {\bf w}_{ij} \quad (i \neq j). &(27)\cr}
$$
Notice that if all terms which vary
inversely with monopole separations are neglected, then the metric
(23) reduces to Bielawski's flat metric.

It can be easily verified that the matrices $g_{ij}$ and
${\bf W}_{ij}$ satisfy
$$
\eqalignno{
{\partial \over \partial x^a_i} W^b_{jk} - {\partial \over \partial
x^b_j} W^a_{ik}&= \epsilon^{abc} {\partial \over \partial x^c_i}
g_{jk},&(28) \cr
{\partial \over \partial x^a_i} g_{jk} &= {\partial \over \partial
x^a_j} g_{ik},&(29)\cr}
$$
where $x^a_i$ and $W^a_{ij}$ denote the components of ${\bf x}_i$ and
${\bf W}_{ij}$.
These conditions, (28) \& (29), were shown by Pedersen and Poon [10], and
Papadopoulos and Townsend [11], following earlier work by
Hitchin et al. [12] to
be the necessary conditions for the metric (23) to be hyperk\"ahler.
These authors gave no explicit solution.  Our explicit solution (24) --
(27) is surprisingly rather simple and symmetric.  In fact, eqns.
(28) \& (29) may be solved locally by a rather more general ansatz than
(24) -- (27).  It suffices to replace $1/r_{ij}$ by a function
$H_{ij} (|{\bf x}_i-{\bf x}_j |)$ which is harmonic in both arguments.
The Dirac potential ${\bf w}_{ij}$ is then chosen to satisfy
${\bf \nabla} _i \times {\bf w}_{ij}= {\bf \nabla}_i H_{ij} \
({\rm no ~ sum ~ over ~} i).$

To obtain the asymptotic region of the $n$-monopole moduli space
$M_n$, one must
quotient $E_n$ by the permutation group $S_n$, which acts by
permuting the positions and phases of the monopoles. This is because the
monopoles are unordered, or indistinguishable. One also requires
$r_{ij} \gg 1$, for all $i \neq j$. Thus in the case of
two monopoles, where the metric calculated using the
method above is the Taub-NUT metric times a flat factor,
one must identify under the interchange of
$({\bf x}_1, \theta_1)$ and $({\bf x}_2, \theta_2)$, and have $r_{12}
\gg 1$, in order to reproduce the asymptotic form of
the moduli space.

The three K\"ahler forms on $E_n$ are
$$
\omega ^a = (d\theta _i + {\bf W}_{i k} \cdot d {\bf x} _k) \wedge
dx^a_i - {1\over 2} g_{ij} \epsilon^{abc}dx^b_i \wedge
dx^c_j, \ \ a=1,2,3. \eqno (30)
$$
It follows that the torus action is triholomorphic
$$
{\cal L}_ {\partial \over \partial \theta _i} \omega ^a = 0 \eqno (31)
$$
and thus that the generators of the torus action, i.e. the Killing
fields $\partial\over \partial \theta_i$, are Hamiltonian vector fields
with respect to all three symplectic forms $\omega^a$.  The three moment
maps associated with $\partial \over \partial \theta_i $ are the three
coordinate functions $x^a_i$.  By contrast the
rotation group $SO(3)$ does not act triholomorphically; the forms
$\omega ^a$ transform as a triplet under its action.

Using the moment maps $x^a_i$ (all of which Poisson commute with
respect to a fixed symplectic structure) one may obtain lower-dimensional
hyperk\"ahler metrics by taking the hyperk\"ahler quotient.
This amounts to setting $k$ of the ${\bf x}_i$'s to constants and then
projecting the metric orthogonally to the $k$ vector fields $\partial
\over \partial {\bf x}_i$.  In this way one obtains a new
$4(n-k)$-dimensional hyperk\"ahler metric.  One may use this freedom
to freeze all but one of
the coordinates ${\bf x}_i$.  The result is a four-dimensional
multi-centre metric of the form discussed in [13]. If one wishes, one
may fix not the ${\bf x}_i$'s but linear combinations of them.  A case of
particular interest is when one reduces using the position
of the centre of mass
$$
{\bf X} = {1\over n} \sum^{n}_{i=1} {\bf x}_i. \eqno (32)
$$
This amounts to taking out the centre of mass motion.  In fact, because
of the special form of the metric (i.e. eqns. (24) -- (27)), it is easy
to see directly that it splits as a  metric product of the
centre of mass manifold, with coordinates ${\bf X}$ and $\theta  =
\sum_i \theta_i$, and a moduli space of centred monopoles.

Let us now turn to the topology of $E_n$.
The space $ \tilde{C}_n ({\Bbb R}^3)$ is the
space of $n$ distinct  ordered points in ${\Bbb R}^3$. It is simply
connected and its second homology group $H_2 (\tilde{C}_n({\Bbb R}^3))$ is
${1 \over 2} n(n-1)$-dimensional and torsion free.
A homology basis is provided by the
${1\over 2}n(n-1)$ $2$-spheres $S_{ij}^{2}~ : ~ |{\bf x}_i - {\bf x}_{j} |
= r_{ij} =$ constant, ${\bf x}_k =$ constant, $j \neq k \neq i$.  To
specify a $T^{n}$ bundle over $\tilde{C}_n ({\Bbb R}^3)$ it suffices
to specify its behaviour over these $2$-spheres.  A $T^n$-invariant
metric gives rise to a connection on the bundle $E_n$ if we define
the horizontal subspaces of the tangent space of $E_n$ to be orthogonal
to the fibre directions with respect to that metric. In our case this
means that the connection  one-form associated to the $i$th generator of
$T^n$ is given by $d\theta _i + {\bf W}_{i k} \cdot d{\bf x}_k$.
Evidently, restricted to $S^2_{ij}$, this reduces to the standard Dirac
monopole connection over $S^2$ with unit magnetic charge.  This
specifies the $T^{n}$ bundle on $\tilde{C}_n ({\Bbb R}^3)$.

In conclusion we have shown that the equations of motion for $n$
well-separated BPS monopoles or dyons can be obtained from the purely kinetic
Lagrangian for motion on a $4n$-dimensional manifold $E_n/S_n$, where $E_n$
is an $n$-torus bundle whose metric (23) is
hyperk\"ahler, and $S_n$ is the permutation group.  The particle
motion in ${\Bbb R^3}$ corresponds to geodesic motion on $E_n/S_n$.
The space $E_n/S_n$ is fibred over
$C_n ({\Bbb R}^3)$, the configuration space of $n$
indistinguishable particles in ${\Bbb
R}^3$. The particles must be at distinct locations, as there are
singularities if they coincide.  $E_n/S_n$ is therefore incomplete. Moreover
if the particles are sufficiently close our metric is no longer
positive definite. This happens because the ``mass'' parameters, i.e.
the coefficients of the second term in (24), are negative.

On the other hand it is known that the low energy dynamics of $n$
BPS monopoles or dyons at arbitrary separation is accurately modelled
by geodesic motion on the moduli space
$M_n$ of static solutions to the Bogomolny equations, and $M_n$ is complete
and thus has no singularities.  Thus $E_n/S_n$ only gives the asymptotic form
of $M_n$, for well-separated monopoles.  In the case
$n=2$, it is known that $E_n/S_n$ and $M_n$ differ by an exponentially small
amount as the separation gets large.  It is natural therefore to conjecture
that $E_n/S_n$  and
$M_n$ are exponentially close, for any $n$, if all pairs of monopoles
are well-separated.  This would imply that in the
scattering of dyons, there would only be exponentially small electric
charge exchange, provided the dyons remained well-separated, although
momentum exchange falls off with a power of the separation.

Mathematically one may consider a version of the metric on $E_n$ for which
the mass parameters are positive. This metric is everywhere positive definite.
For suitably chosen values of the mass parameters it is also complete,
as the subspace where pairs of points in ${\Bbb R}^3$ coincide
corresponds to the fixed point
set of one of the generators of the torus action. The simplest case
is $n=2$ which gives the usual Taub-NUT metric with positive mass. We
believe that the global structure of these positive definite metrics
would repay further study in view
of their application as target spaces for supersymmetric sigma-models.

We have made no progress with solving the equations of motion for $n$
well-separated dyons.  In the $n=2$  case, we found enough constants
of motion to determine the asymptotic $2$-dyon scattering and bound orbits
explicitly [14].  It would be very interesting if the $n$-body problem
was tractable. We have however found a simple expression for the
total angular momentum which includes the
contribution from the interaction of electric and magnetic charges, and
which generalizes that found in [14]. It is

$$
{\bf J}= 4\pi \sum _{i,j} g_{ij} {\bf x}_i \times {\bf v}_j -
\sum _{i<j} (q_i-q_j) {{\bf r}_{ij}\over r_{ij} } , \eqno (33)
$$
or equivalently,
$$
{\bf J}= 4\pi \sum _{i=1}^n {\bf x}_i \times {\bf v}_i -
\sum _{1 \leq i<j \leq n} {4\pi ({\bf x}_i - {\bf x}_j) \times ({\bf
v}_i - {\bf v}_j) + ({\bf x}_i - {\bf x}_j)(q_i-q_j) \over r_{ij} } .
\eqno (34)
$$

It is interesting to study the quantized dynamics of monopoles, and
especially the dynamics of monopoles in super-Yang-Mills theory.  Sen
has recently argued that there are certain quantum bound states of $n$
monopoles with zero energy in such a theory [15].  To verify Sen's
conjectures one first has to understand the cohomology of the
$n$-monopole moduli space $M_n$, which depends largely on the
properties of $M_n$ when all monopoles are close together.  However
to check in detail whether the cohomology classes are directly related
to normalizable quantum states requires a careful analysis of the
metric on $M_n$ for well-separated monopoles, and here our results may
be useful.  In the simplest case of two monopoles, the bound state
wavefunction decays exponentially with separation and it may be that
for $n$ monopoles, the wavefunction similarly decays if any one
monopole becomes separated from the rest.

Finally we remark that if we set the electric charges to zero
in (13) we obtain a metric on $({\Bbb R}^3)^n$ which is up to rescaling
almost identical to that obtained by Shiraishi [16] for $n$ stringy
black holes.  The difference is that in Shiraishi's case the minus sign
between the first two terms in (13) is replaced by a plus sign.  This
would seem to imply that by adjoining extra angles to the moduli space
of $n$ stringy black holes one may obtain a hyperk\"ahler moduli
space.  This may lead to a resolution of the puzzle discussed in [17].

\vskip 30pt

{\bf References}

\item {[1]} M.F. Atiyah and N.J. Hitchin, The geometry and dynamics of
magnetic monopoles (Princeton University Press, 1988).

\item {[2]} N.J. Hitchin, N.S. Manton and M.K. Murray, Symmetric
Monopoles (DAMTP preprint 95-17, 1995).

\item {[3]} N.S. Manton, Phys. Lett. B 110 (1982) 54.

\item {[4]} D. Stuart, Commun. Math. Phys. 166 (1994) 149.

\item {[5]} R. Bielawski, Monopoles, particles and rational
functions (McMaster University preprint, 1994).

\item {[6]} M.F. Atiyah and N.J. Hitchin, Phys. Lett. A 107 (1985) 21;
Phil. Trans. R. Soc. Lond. A 315 (1985) 459.

\item {[7]} S.W. Hawking, Phys. Lett. A 60 (1977) 81.

\item {[8]} N.S. Manton, Phys. Lett. B 154 (1985) 397; (E) B 157
(1985) 475.

\item {[9]} B. Julia and A. Zee, Phys. Rev. D 11 (1975) 2727.

\item {[10]} H. Pedersen and Y.S. Poon, Commun. Math. Phys. 117 (1988) 569.

\item {[11]} G. Papadopoulos and P.K. Townsend, Nucl. Phys. B 444
(1995) 245.

\item {[12]} N.J. Hitchin, A. Karlhede, U. Lindstr\"om and M. Ro\v cek,
Commun. Math. Phys. 108 (1987) 537.

\item {[13]} G.W. Gibbons and S.W. Hawking, Phys. Lett. B 78 (1978) 430.

\item {[14]} G.W. Gibbons and N.S. Manton, Nucl. Phys. B 274 (1986) 183.

\item {[15]} A. Sen, Phys. Lett. B 329 (1994) 217.

\item {[16]} K. Shiraishi, Nucl. Phys. B 402 (1993) 399.

\item {[17]} G.W Gibbons and R. Kallosh, Phys. Rev. D 51 (1995) 2839.

\bye